\documentstyle[aps,epsfig,twocolumn,floats,array,multirow]{revtex}

\title{Optimization of evaporative cooling towards a large 
number of Bose-Einstein condensed atoms}
\bigskip
\author{Makoto Yamashita,$^1$ Masato Koashi,$^2$ Tetsuya Mukai,$^1$ 
Masaharu Mitsunaga,$^3$ \\
Nobuyuki Imoto,$^{1,2}$ and Takaaki Mukai$^1$}
\bigskip
\address{
$^1$NTT Basic Research Laboratories, NTT Corporation, 
3-1, Morinosato Wakamiya, Atsugi-shi, Kanagawa 243-0198, Japan 
}
\address{
$^2$The Graduate University for Advanced Studies, 
Shonan Village, Hayama, Kanagawa 240-0193, Japan
}
\address{
$^3$Kumamoto University, 
Faculty of Science, Department of Physics, 2-39-1, Kurokami, 
Kumamoto 860-8555, Japan
}
\date{\today}
\begin{document}
\maketitle
\begin{abstract}
We study the optimization of evaporative cooling in 
trapped bosonic atoms on the basis of quantum kinetic theory of a Bose gas. 
The optimized cooling trajectory for $^{87}$Rb atoms indicates that 
the acceleration of evaporative cooling around the transition point of 
Bose-Einstein condensation is very effective against loss 
of trapped atoms caused by three-body recombination. 
The number of condensed atoms is largely enhanced 
by the optimization,   
more than two orders of magnitude in our present 
calculation using relevant experimental parameters, 
as compared with the typical value given by the 
conventional evaporative 
cooling where the frequency of radio-frequency magnetic 
field is swept exponentially. 
In addition to this optimized cooling, it is also shown that 
highly efficient evaporative cooling can be achieved by an 
initial exponential and then a rapid linear sweep of frequency.   
\end{abstract}\bigskip
Pacs. 03.75.Fi, 05.20.Dd, 32.80.Pj
\vspace{1.0cm}
\section{Introduction}
Evaporative cooling is an essential experimental technique in 
the recent demonstrations of Bose-Einstein condensation 
(BEC) with magnetically trapped atomic vapor \cite{Rb,Li,Na,H,He}.  
This cooling method is based on both the selective removal of energetic 
atoms through evaporation and collisional rethermalizations among 
remaining atoms \cite{eva,Luit}. 
The successful utilization of interatomic elastic collisions makes 
evaporative cooling highly efficient. A real system, however, involves other 
atomic collisions, such as elastic collisions with background gas, 
inelastic collisions due to dipolar relaxation and three-body 
recombination processes.  
These undesirable collisions result in a loss of the trapped atoms 
and seriously reduce the effectiveness of evaporative cooling.  
\par
For alkali-metal atoms, such as $^{87}$Rb  
and $^{23}$Na, three-body recombination becomes a dominant loss 
mechanism during evaporative cooling \cite{eva}, and limits the final 
number of condensed atoms, $N_0$, in the experiments.  
It is known that $^{87}$Rb atoms suffer much severer losses than 
$^{23}$Na atoms due to their much large rate coefficient of three-body 
recombination process \cite{Rb_3body,Na_3body}. 
BEC experiments have clearly reflected this fact: 
the reported $N_0$ value in $^{87}$Rb is $10^6$ at 
most \cite{Rbmax}, while that in $^{23}$Na is up to $5 \times 10^7$ 
\cite{Namax}.  
The small BEC of $^{87}$Rb is a great disadvantage with respect to 
its future application as an ^^ ^^ atom laser"\cite{AtomLaser} and  
has to be overcome. 
\par
In this paper, we report the theoretical optimization of evaporative cooling,  
focusing on BEC experiments with $^{87}$Rb atoms. 
Several theoretical works have considered the optimization of 
evaporative cooling to increase the cooling efficiency \cite{Sackett,1D_ev,eva}. 
Especially, Ref.\ \cite{1D_ev} optimized the evaporative cooling in atomic hydrogen  
not only theoretically but also experimentally. 
However, these studies are based on kinetic theory of a classical 
gas which fails around the BEC transition point, 
and therefore  don't give us any information on the number of condensed atoms 
produced by evaporative cooling. 
Our analysis is based on quantum kinetic 
theory of a Bose gas \cite{Yama} and can provide quantitative results on the increase in   
the number of condensed atoms by optimizing the evaporative cooling. 
We optimize the temporal frequency sweep of the radio-frequency (rf) magnetic 
field under a fixed trapping potential by following the variational method 
developed by Sackett {\it et al.} for evaporative 
cooling in a classical gas system \cite{Sackett}.  
We show that an accelerated sweep of the rf-field frequency 
around the BEC transition point can largely enhance the final number of condensed 
atoms compared with conventional cooling 
with an exponential sweep of the frequency. 
It is also demonstrated that 
highly efficient evaporative cooling can be achieved by an 
initial exponential and then a rapid linear sweep of frequency.   
Our present results provide a useful guideline to experimentalists aiming to 
achieve large condensates in $^{87}$Rb atoms. 
\par
\bigskip 
\section{Kinetic theory of evaporative cooling}
\subsection{Thermodynamics of trapped bosonic atoms}
We begin with a review of our theoretical formulation \cite{Yama}. 
During evaporative cooling, a magnetic potential, $U({\bf r})$, is 
truncated depending on the frequency of 
the applied rf-magnetic field. 
The thermalized distribution of 
noncondensed atoms in such a truncated potential 
is well approximated by the truncated Bose-Einstein distribution function such that  
\begin{equation}
\tilde f( {\bf r}, p) = \frac{1}{\exp(\epsilon_p/k_BT)/\tilde \xi({\bf r})-1} \ 
\Theta(\tilde A({\bf r}) -\epsilon_p),
\end{equation}
where $T$ is the temperature, $\epsilon_p=p^2/2m$ is the kinetic energy of 
atoms with momentum $p$ and 
mass $m$, and $\Theta(x)$ is a step function. 
In this paper, we add tilde to the notations of the quantities evaluated through this 
truncated distribution function. 
$\tilde \xi({\bf r})$ denotes the local fugacity including mean-field potential energy; 
$\tilde \xi ({\bf r})=\exp\{[\mu-U({\bf r})-2v \tilde n ({\bf r})]/k_BT\}$, where 
$\mu $ is the chemical potential, $v=4 \pi  a \hbar^2/m$ is the interaction 
strength of atoms in proportional to the {\it s}-wave scattering length $a$,
and $\tilde n({\bf r})$ is the density profile of atoms. 
The step function eliminates 
the momentum states whose kinetic energy exceeds the effective potential height 
$\tilde A({\bf r})=\epsilon_t-U({\bf r})-2v\tilde n({\bf r})$, where $\epsilon_t$ is the 
truncation energy of a magnetic potential. 
We calculate the density profile of atoms using this truncated distribution function 
in a self-consistent way  such that 
\bigskip
\begin{equation}
\tilde n({\bf r})=4 \pi h^{-3} \int \tilde f({\bf r},p) p^2dp, 
\end{equation}
and the internal energy density similarly as 
\bigskip
\begin{equation}
\tilde e({\bf r})=4 \pi h^{-3} \int \epsilon_p\tilde f({\bf r},p) p^2dp + 
v {\tilde n}^2({\bf r})+U({\bf r})\tilde n({\bf r}).
\end{equation}
The spatial integrations of these density functions give the total number of atoms, 
$\tilde N=\int\tilde n({\bf r}) d{\bf r}$, and the total internal energy, 
$\tilde E=\int\tilde e({\bf r}) d{\bf r}$ respectively.  
\par
After the BEC transition, the atom density in the  condensed region is obtained by 
the sum of a condensate $n_0({\bf r})$ 
and  saturated noncondensed atoms $\tilde n_n$ such that 
$\tilde n({\bf r})=n_0({\bf r})+\tilde n_n$. 
The condensate is described by the Thomas-Fermi distribution $n_0({\bf r})=
n_p-U({\bf r})/v$ with the peak density of the condensate at the center of potential 
$n_p$, while $\tilde n_n$ is evaluated through the truncated  Bose-Einstein 
distribution function with the condition $\tilde \xi({\bf r})=1$. 
\par
\bigskip
\subsection{Dynamics of evaporative cooling}
The dynamics of trapped atoms during evaporative cooling is investigated 
on the basis of quantum kinetic theory of a Bose gas, developed previously by 
the present authors \cite{Yama}. Trapped atoms are removed 
from the potential 
by not only evaporation but also undesirable collisions, as mentioned in the 
introduction. 
The change rates (i.e., loss rates) of the total number of atoms and the total 
internal energy are calculated respectively as the 
sum of the contributions of all processes such that,  
\bigskip
\begin{eqnarray}
\label{difeq}
\nonumber  \frac{d \tilde N}{dt} &=&-\int\ \dot n_{\rm ev}({\bf r})\ {\rm d}{\bf r} \\
\nonumber  & &-
\sum_{s=1}^3 \ G_s \int K_s({\bf r}) [\tilde n({\bf r}) ]^s\ {\rm d}{\bf r}
+\left( \frac{\partial \tilde N }{\partial \epsilon_t}\right)_{T, \mu} \dot \epsilon_t, \\
\frac{d \tilde E}{dt} &=&-\int\ \dot e_{\rm ev} ({\bf r})\ {\rm d}{\bf r} \\
& &
-\sum_{s=1}^3 
\nonumber  G_s \int K_s({\bf r}) \tilde e({\bf r}) [\tilde n({\bf r})]^{s-1}\ {\rm d}{\bf r} 
+\left( \frac{\partial \tilde E }{\partial \epsilon_t}\right)_{T, \mu} \dot \epsilon_t.
\end{eqnarray}  
The rates $\dot n_{\rm ev}$ and $\dot e_{\rm ev}$ denote the evaporation rates of 
density functions derived from a general collision integral of 
a Bose gas system \cite{Yama,eva_rate}.  
The parameters $G_1$,  $G_2$, and $G_3$ are the decay rate constants of 
trapped atoms due to the background gas collisions, dipolar relaxation, 
and three-body recombination, respectively. 
$K_s$ represents the correlation function which describes the $s$-th 
order coherence of trapped atoms, and we assume the 
expressions of $K_s$ for an ideal Bose 
gas system \cite{kagan,Rb_3body}. 
The terms proportional to the change rate of truncation energy, 
$\dot \epsilon_t=d\epsilon_t/dt$, give the contribution of extra atoms 
^^ ^^ spilled over" when $\epsilon_t$ changes continuously in the forced 
evaporative cooling \cite{Kristine,1D_ev}. 
The dynamics of the evaporative cooling including the loss of trapped atoms is 
described by the coupled differential equations in Eq.\ (\ref{difeq}). 
In our theoretical framework based on truncated 
Bose-Einstein distribution function, all thermodynamic quantities 
are given as the complicated functions of three independent variables: 
temperature $T$, chemical potential $\mu$, and truncation 
energy $\epsilon_t$. 
We derive here the 
useful thermodynamic relations in terms of   
the change rates of these variables: 
\bigskip
\begin{eqnarray}
\nonumber \displaystyle \frac{d \tilde N}{dt} &=& 
\left( \frac{\partial \tilde N }{\partial T}\right)_{\mu, \epsilon_t} \dot T
+ \left( \frac{\partial \tilde N }{\partial \mu}\right)_{T, \epsilon_t} \dot \mu 
+\left( \frac{\partial \tilde N }{\partial \epsilon_t}\right)_{T, \mu} \dot \epsilon_t,  \\
\displaystyle \frac{d \tilde E}{d t} &= &
\left( \frac{\partial \tilde E }{\partial T}\right)_{\mu, \epsilon_t} \dot T
+ \left( \frac{\partial \tilde E }{\partial \mu}\right)_{T, \epsilon_t} \dot \mu 
+\left( \frac{\partial \tilde E }{\partial \epsilon_t}\right)_{T, \mu} \dot \epsilon_t, 
\label{tempdif}
\end{eqnarray}
where $\dot T$ and $\dot \mu$ are the change rate of temperature and that of chemical 
potential respectively. 
Substituting these relations into Eq.\ (\ref{difeq}), we clearly see that the terms 
proportional to $\dot \epsilon_t$ are canceled out and both $\dot T$ and $\dot \mu$ are 
determined by three variables $T$, $\mu$, and $\epsilon_t$. 
\par
\bigskip
\subsection{Optimization of cooling trajectory}
The optimum cooling trajectory in the evaporative cooling process 
is obtained approximately by following the instantaneous optimization procedure 
investigated in Ref. \cite{Sackett}.  
We determine the optimized cooling trajectory so as to achieve 
the maximum increase in phase-space density $\tilde \rho$ with the 
smallest decrease in the number of trapped atoms $\tilde N$. 
In the nonuniform gas system, the phase-space density is evaluated at the peak 
position and given by $\tilde \rho=\tilde n({\bf 0}) \lambda^3$ with the 
thermal de Broglie wavelength $\lambda=h/\sqrt{2 \pi m k_BT}$. 
At each time, we calculate the optimum truncation energy $\epsilon_t^\ast$  
by maximizing the derivative, $-d( \ln \tilde \rho)/d( \ln \tilde N$), 
which describes the efficiency of evaporative cooling 
\cite{Sackett,eva}. 
We use a more useful form of this derivative rewritten 
in terms of change rates of temperature, chemical potential, and truncation energy as, 
\begin{eqnarray}
\displaystyle  
\nonumber - \frac{d (\ln \tilde \rho) }{ d (\ln \tilde N )} &=& 
 - \frac{\left( \frac{\partial \tilde \rho}
{\partial T}\right )_{\mu, \epsilon_t} \dot
T+\left( \frac{\partial \tilde \rho}{\partial \mu}\right)_{T,\epsilon_t} \dot \mu 
+\left( \frac{\partial \tilde \rho }{\partial\epsilon_t}\right)_{T, \mu} \dot \epsilon_t}
{\left(\frac{\partial \tilde N}{\partial T}\right)_{\mu,\epsilon_t} \dot
T+\left( \frac{\partial \tilde N}{\partial \mu} \right)_{T,\epsilon_t} \dot \mu
+\left( \frac{\partial \tilde N }{\partial
\epsilon_t}\right)_{T, \mu} \dot \epsilon_t }\\
& & \times 
\left(\frac{\tilde N}{\tilde \rho} \right). 
\label{derivative}
\end{eqnarray}
In the optimized evaporative cooling, the truncation parameter $\eta=\epsilon_t/k_BT$ 
changes more slowly than the variables $T$, $\mu$, and $\epsilon_t$. 
We neglect the change rate of this truncation parameter, 
and consequently $\dot \epsilon_t$ is related to $\dot T$ as 
$\dot \epsilon_t/\epsilon_t=\dot T/T$ 
in Eq.\ (\ref{derivative}). 
The value of optimum truncation energy $\epsilon_t^\ast$ is given as a 
function of both temperature $T$ and chemical potential $\mu$. 
Substituting this $\epsilon_t^\ast$ into Eq.\ (\ref{difeq}), we obtain 
the coupled differential equations describing the dynamics of 
the optimized evaporative cooling.  
The time-evolution calculations are carried out numerically by 
repeating a series of steps consisting of instantaneous optimization and 
dynamical short-time evolution of Eq.\ (\ref{difeq}) with the temporal value 
of $\epsilon_t^\ast$. 
Furthermore we assume quick rethermalization, 
where the system is always described by the truncated distribution 
function \cite{Yama,Luit}.  
The obtained function $\epsilon_t^{\ast}(t)$ 
corresponds to the optimized sweeping function of the rf-field frequency. 
\par 
In the next section, we will show the numerical results calculated by using relevant 
experimental parameters and discuss the experimental possibility of making large condensates 
in $^{87}$Rb atoms. 
\par
\bigskip
\section{Numerical Results and Discussions}
\subsection{Optimized evaporative cooling}
In our numerical calculations, we employed parameters corresponding to 
the experiments performed at the University of Tokyo \cite{Torii}. 
The cloverleaf magnetic trap with bias field $B_0$, radial gradient $B^{'}$, and axial 
curvature  $B^{''}$ was modeled by
the magnetic field of Ioffe-Pritchard type such that 
$B(r,z)=\sqrt{(\alpha r)^2+(\beta z^2+B_0)^2}$,  
where $\alpha=\sqrt{B^{'2}-B^{''}B_0/2}$  
and $\beta=B^{''}/2$. We used the typical values adopted experimentally,  
such as $B_0=0.9\ {\rm G}$, $B^{'}=180\ {\rm G/cm}$,  and $B^{''}=170\ {\rm G/cm^2}$. 
The corresponding trapping frequencies are $\omega_r=2 \pi \times 171$\ Hz in 
the radial direction, 
and $\omega_z=2 \pi \times 11.8$\ Hz in the axial direction. 
The collisional parameters for the trapped ${\rm F}=1, {\rm m_F} =-1$ state 
were set as follows \cite{Rb_3body}: {\it s}-wave scattering length
$a=5.8$ nm,  loss rate constants due to background gas collisions 
$G_1=0.01 \ {\rm s^{-1}}$, dipolar relaxation $G_2=1.5\times10^{-16}\ 
{\rm cm^3 s^{-1}}$, and three-body recombination
$G_3=4.3\times10^{-29}\ {\rm cm^6 s^{-1}}$.   
We chose the temperature 
$T=150 \ \mu$K, the peak density $\tilde n ({\bf 0})=4.0\times 10^{11}$ cm$^{-3}$, 
and the number of trapped atoms $ \tilde N =1.0\times 10^{9}$ as the initial 
evaporative cooling conditions. 
\par
The exponential sweep of the rf-field frequency is usually adopted in BEC 
experiments \cite{Torii}. 
We also performed calculations assuming that the frequency, $\nu$, is swept 
exponentially from 35 to 0.67 MHz in 60 s as in Ref.\cite{Torii}. 
The comparison between the different sweeps of rf-field frequency, optimized or 
exponential, will clarify the merit of the present optimization procedure. 
\par
\begin{figure}[htb]
  \begin{center}
    \leavevmode
 \begin{minipage}{8.0cm}
    \epsfig{file=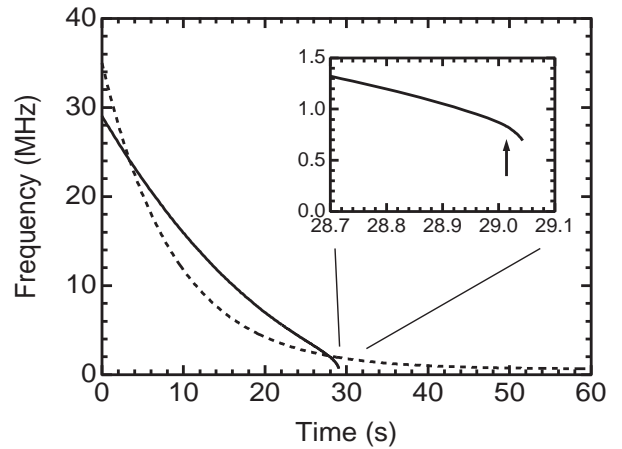,angle=0,width=\linewidth}
 \end{minipage}
  \end{center}
\caption{
Sweeps of rf-field frequency in the optimized case (solid) and 
in the exponential case (dashed). The inset shows the last part of the 
optimized sweep using an expanded scale. The arrow in the inset 
indicates the point at which the BEC transition occurs.}
 \label{optifig}
\end{figure}
\bigskip
Figure 1 shows each sweep of the rf-field frequency.  We see that the 
sweeping time (i.e., the cooling time) in the optimized case is about 
half that of exponential-sweep cooling for 60 s. The shorter sweeping time in 
the optimized cooling is advantageous for BEC experiments since the long time 
evaporative cooling easily becomes inefficient due to the instability of 
the magnetic trap. 
Furthermore, it is found that the optimized frequency sweep becomes rapid at the 
final stage of the cooling process, i.e., after about 27 s, and such a sweep is quite 
opposite to the exponential one which slows down over time. 
As shown in the inset of Fig.\ 1, the optimization procedure accelerates the 
evaporation around the BEC transition point.  Similar acceleration of 
evaporation at the final stage of optimized cooling has been also demonstrated 
by the previous studies based on kinetic theory of a classical gas \cite{Sackett,1D_ev}. 
\par
\begin{figure}[htb]
  \begin{center}
    \leavevmode
 \begin{minipage}{8.0cm}
    \epsfig{file=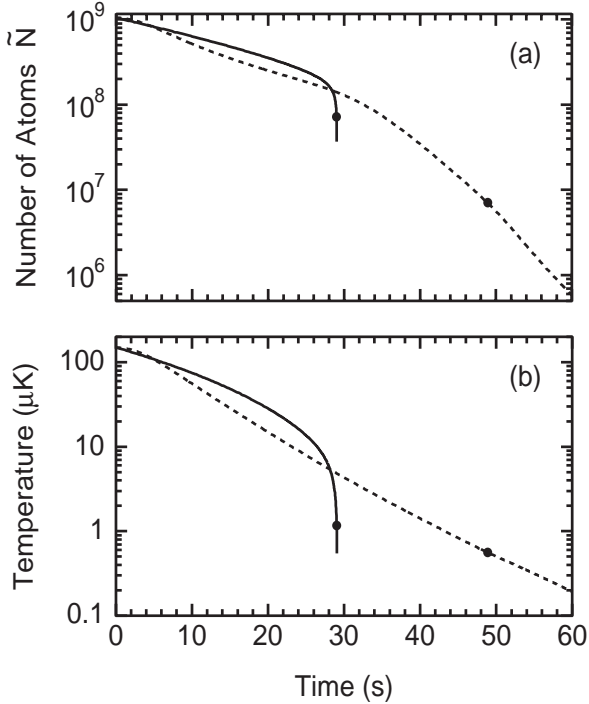,angle=0,width=\linewidth}
 \end{minipage}
  \end{center}
\caption{
Time evolution of (a) the total number of trapped atoms $\tilde N$ 
and (b) temperature $T$, for optimized cooling 
(solid) and for exponential-sweep cooling (dashed). 
The dots indicate the points at which the BEC transition occurs. }
 \label{opti_N_T}
\end{figure}
\bigskip
In Figs.\ 2(a) and 2(b), we show the time evolution of the number of trapped 
atoms $\tilde N$ and that of temperature $T$, respectively.  
In the optimized cooling, both  $\tilde N$ and $T$ rapidly decrease at the 
final stage of cooling due to the rapid evaporation mentioned above. 
On the other hand, 
$\tilde N$ and $T$ decrease monotonically in the exponential cooling case. 
At the BEC transition point indicated by the dots in the figure, 
both the number of trapped atoms and temperature are calculated to be 
$\tilde N=7.2\times 10^7$ and  $T=1.2 \ \mu$K at 29.01 s for the 
optimized cooling, and $\tilde N=7.1 \times 10^6$ and $T=0.56\ \mu$K 
at 48.9 s for the exponential cooling. 
\par
\begin{figure}[htb]
  \begin{center}
    \leavevmode
 \begin{minipage}{8.0cm}
    \epsfig{file=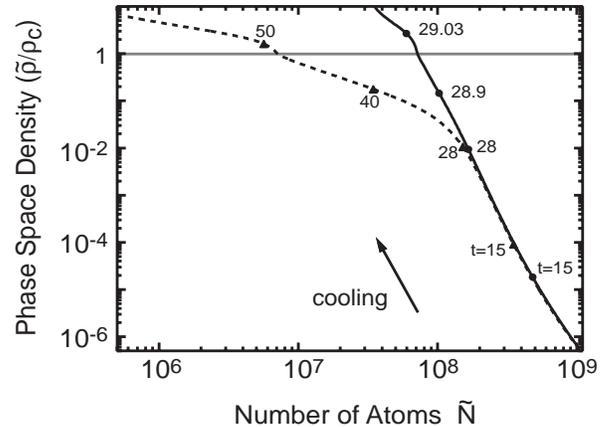,angle=0,width=\linewidth}
 \end{minipage}
  \end{center}
\caption{
Cooling trajectory through phase-space for optimized cooling 
(solid) and for exponential-sweep cooling (dashed). 
The phase-space 
density $\tilde \rho $ is normalized by the 
critical value $\rho_c=2.612$. The BEC transition occurs when each 
trajectory passes the horizontal line where $\tilde \rho/\rho_c=1$. 
The arrow indicates the direction of time. Several corresponding time 
values are added to the trajectories to demonstrate the time evolution 
of the system.}
 \label{trajectory}
\end{figure}
\bigskip
Next, we show how the optimization affects the cooling efficiency in the whole 
evaporative cooling process. 
In Fig.\ 3, the cooling trajectories of both cases are plotted in the
$\tilde N$-$\tilde \rho$ plane and several corresponding time values are also 
added to the cooling trajectories to demonstrate the time evolution.  
At the early stage of cooling where the 
loss due to background gas collisions is dominant, both trajectories 
are quite similar,  indicating that the exponential sweep of rf-field 
frequency is sufficiently good. 
However, the difference becomes larger as the trajectories approach 
the quantum degenerate region where the phase-space density exceeds 
the critical value of $\rho_c=2.612$ and Bose statistics of atoms becomes 
dominant.   
Clearly, more atoms are lost in conventional exponential-sweep cooling 
than in the optimized case. 
For example, we have ten times larger number of atoms in the optimized cooling 
at the BEC transition point when the trajectory reaches the horizontal line 
of $\tilde \rho/\rho_c=1$.  
Note that the point where the two trajectories split up 
in Fig.\ 3 (i.e., $\tilde \rho/\rho_c \simeq 0.01$) corresponds to the time 
of around 28 s where the rf-field frequency curves cross in Fig.\ 1. 
The sudden increase in phase-space density just after the BEC transition 
in both trajectories reflects the rapid growth of condensate due to 
bosonic stimulation. 
\par
The number of condensed atoms, 
$N_0$, is largely increased by the optimization.  In the optimized cooling, $N_0$  
increased monotonically and we obtained $N_0=2.1\times 10^7$ after 
29.04-s cooling. In the exponential-sweep cooling, on the other hand, 
$N_0$ had the maximum value of $1.74\times 10^5$ at 57.0 s and then 
gradually decreased. 
We finally obtained  $N_0=1.65 \times 10^5$ for 60-s exponential sweep cooling.  
More than a 100-fold enhancement of the number of condensed atoms was achieved 
in our present calculations.  
The lifetime of such a large condensate is evaluated to be less than 100 ms due to
the enormous three-body recombination loss, as we will see in Fig.\ 4. 
However, it is possible to resolve this problem by adiabatically expanding the 
trapping potential. We also add that the value of $N_0=1.7\times 10^5$ quantitatively 
agrees with the experimental results reported in Ref.\cite{Torii} and confirms 
the validity of our numerical calculations. 
\begin{figure}[htb]
  \begin{center}
    \leavevmode
 \begin{minipage}{8.2cm}
    \epsfig{file=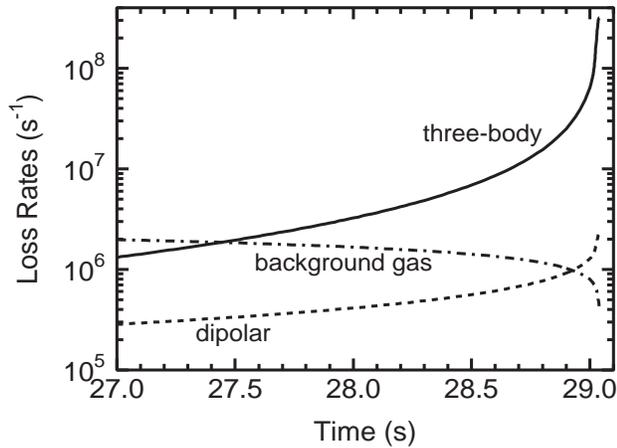,angle=0,width=\linewidth}
 \end{minipage}
  \end{center}
\caption{
Time evolution of loss rates at the final stage of 
optimized evaporative cooling, loss due to the background gas 
collision (dot-dashed), the dipolar relaxation process (dashed), 
and the three-body recombination process (solid).}
 \label{loss_rate}
\end{figure}
\par
The undesirable loss of trapped atoms during the cooling process gives us the 
physical origin of the rapid frequency sweep in the optimized cooling. 
Figure 4 shows the time evolution of loss rates of total number of trapped atoms 
at the final stage of the optimized cooling. 
It is seen that the loss due to the three-body
recombination greatly increases and selectively removes atoms with 
high density (i.e., atoms with lower energy).  
This means that the serious heating of the system is caused by 
three-body recombination loss.  
In order to get a large number of condensed atoms, 
evaporative cooling should therefore be accelerated to 
compensate for such undesirable heating. 
\par 
Here, we briefly discuss the validity of the quick rethermalization assumption 
in the rapid evaporation necessary for optimized cooling. 
The evolution of the system in the evaporative cooling process is well scaled 
by the characteristic time 
$\tau$ given by the inverse of normalized loss rate as 
$\tau = (-\dot N_{\rm loss}/\tilde N)^{-1}$. 
The thermalization of the system, on the other hand, is characterized 
by the elastic collision time 
$\tau_{\rm coll}=(n \sigma v )^{-1}$; 
$n$ is the mean atom density, $\sigma=8 \pi a^2$ is the elastic 
collision cross section, 
and $v = 4\sqrt{k_B T/ \pi m }$ is the mean velocity of 
relative motion of the atoms \cite{Wu}. At the BEC transition point in 
the optimized cooling, 
these characteristic times are calculated to be 
$\tau = 140$ ms and $\tau_{\rm coll}=0.4$ ms, respectively. 
The relation $\tau \gg \tau_{\rm coll}$ confirms that quick rethermalization
is still valid during the rapid evaporation that occurs for optimized cooling.  
\par
\bigskip
\subsection{Efficient evaporative cooling based on general sweeping functions of rf-frequency}
The optimized frequency sweep in Fig.\ 1  
is the complicated function of the experimental parameters such as 
the temperature, the density of gas, the atomic collisional parameters, 
the magnetic trapping potential, and so on. This means that we need to 
calculate the optimized cooling for each BEC experiment  
depending on its experimental conditions. 
Here we try to demonstrate the efficient 
evaporative cooling based on general sweeping functions of rf-field frequency, 
which is useful for many experimentalists who want to achieve large condensates. 
\par
\begin{figure}[h]
  \begin{center}
   \leavevmode
 \begin{minipage}{7.5cm}
    \epsfig{file=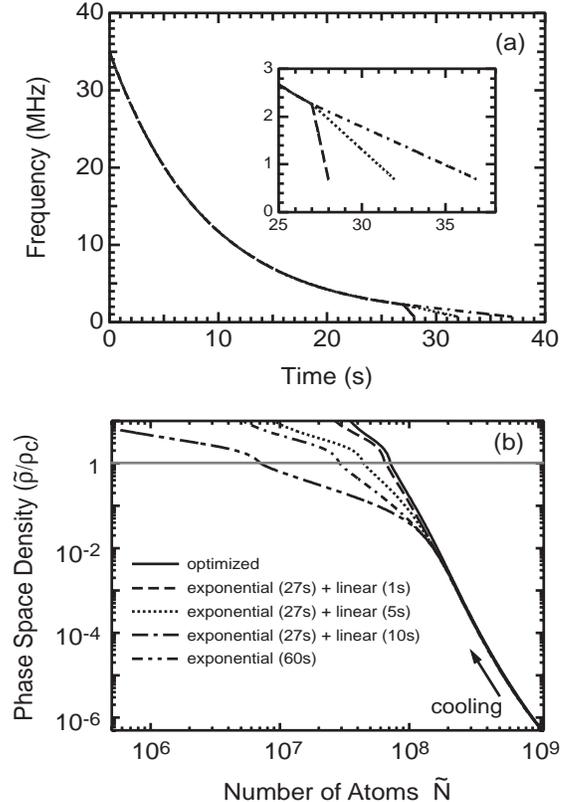,angle=0,width=\linewidth}
 \end{minipage}
  \end{center}
\caption{
(a) Sweeps of rf-field frequency consisting of two segments: 
an identical exponential sweep for 27 s 
and a sequential linear sweep for three different times which  
are 1, 5, and 10 s respectively. 
The exponential sweep is equal to the corresponding part of the 
dashed line in Fig.\ 1. 
The inset shows the linear-sweep region with an expanded scale. 
(b) Cooling trajectories through phase-space for the 
evaporative cooling using the sweeping functions in (a). 
The results of both optimized cooling and exponential 
cooling in Fig.\ 3 are also plotted for comparison. 
The phase-space density $\tilde \rho $ is normalized by the 
critical value $\rho_c=2.612$. The BEC transition occurs when each 
trajectory passes the horizontal line where $\tilde \rho/\rho_c=1$.
The arrow indicates the direction of time. }
 \label{exp_lin}
\end{figure}
\bigskip
From the results in Figs.\ 3 and 4, the exponential sweep is highly efficient until 
the three-body recombination loss becomes dominant. 
The rapid frequency sweep at the final stage of the optimized cooling in Fig.\ 1 
might be approximated by a linear function. 
Thus it is expected that a combination of an exponential sweep of rf frequency 
and a rapid linear one realizes the high cooling efficiency. 
To test this cooling method, we carried out the time-evolution calculation by applying 
the same numerical parameters used in the optimization calculation 
of the proceeding subsection. 
As shown in Fig.\ 5(a), the frequency is first swept exponentially from 35 MHz to 
2.26 MHz for 27 s and then swept linearly to 0.67 MHz for three different times 
of 1 s, 5 s, and 10 s respectively.  
The exponential curve in Fig.\ 5(a) is in conformity with the corresponding part of 
the dashed line in Fig.\ 1. 
\par
Figure 5(b) shows the cooling trajectories in the $\tilde N$-$\tilde \rho$ plane for the 
evaporative cooling using the sweeping functions of Fig.\ 5(a), 
together with those for the optimized and exponential 
cooling plotted in Fig.\ 3. We can see that the linear frequency sweep at the final stage of 
the evaporative cooling strongly improves the cooling efficiency as expected. 
Especially, the cooling trajectory in the case of 27-s exponential and 1-s linear frequency 
sweep (dashed line) is very close to the optimized one (solid line). 
The final number of condensed atoms in this cooling method 
is calculated to be $N_0=1.8\times 10^7$ and comparable to the value in the optimized cooling 
such as $N_0=2.1 \times 10^7$. 
One can achieve the evaporative cooling with sufficiently high efficiency 
by an initial exponential and a sequential rapid linear sweep of rf-field frequency. 
\par
\subsection{Experimental realization of efficient evaporative cooling}
Recently, a group at Gakushuin University performed the efficient evaporative cooling 
for $^{87}$Rb atoms with F=2, ${\rm m}_{\rm F}=2$ trapped state \cite{Gakushuin}. 
They applied a frequency sweep similar to the optimized one which
we  calculated according to their experimental situations.  
The time of evaporative cooling was  divided into three segments 
and in each segment the optimized sweeping function was approximated by the linear one:  
the frequency was swept linearly like 32 $\rightarrow$13 $\rightarrow$ 2 
$\rightarrow $1.1 MHz for 10, 10, and 5 s respectively. 
A large number of condensed atoms,  $N_0=1.4 \times 10^6$, were finally obtained 
after this 25-s evaporative cooling starting from the experimental 
conditions such as initial temperature $T=380 \ \mu K$, 
initial total number of atoms $N=4\times 10^8$, 
and initial phase-space density $\rho=5.8\times 10^{-8}$. \par
On the other hand, we theoretically investigated this evaporative cooling. 
It was found that the cooling trajectory through phase space 
is close to that of optimized cooling and the final 5-s linear sweep 
of rf-frequency enhances the phase space density more than four orders of magnitude.  
The number of condensed atoms for this 25-s evaporative cooling 
was calculated to be  $1.7 \times 10^6$. 
Our calculated results agree well with the experiment at Gakushuin University. 
\section{Conclusion}
We have theoretically studied the optimization of evaporative cooling on the basis of 
quantum kinetic theory of a Bose gas,  
focusing on BEC experiments with rubidium atoms. The calculated results indicate 
that the acceleration of evaporative cooling around the BEC transition point is very 
effective against serious three-body recombination loss. 
The number of condensed atoms is expected to be largely increased by this 
optimization procedure. Furthermore, 
it is shown that highly efficient evaporative cooling can be achieved 
by an initial exponential and then a rapid linear sweep of frequency.   
These results are consistent with the experiments and provide a useful guideline to 
experimentalists aiming to achieve large condensates in $^{87}$Rb atoms. 
\par
In the present work, we optimized the sweeping function of 
the rf-field frequency only. 
A more complete optimization together with 
the adiabatic change of trapping potential curvature 
might produce a large and long-lived condensate. 
On the other hand, our theory neglects the influence of gravity which 
reduces the cooling efficiency at low temperatures \cite{eva}. 
An interesting future work would be to include the gravity in 
our present kinetic analysis and study how the optimized cooling 
trajectory is modified by it. 
\par
\section*{Acknowledgments}
We are grateful to Y. Yoshikawa,  K. Araki, T. Kuwamoto, T. Hirano of Gakushuin University, 
M. Kozuma of the Tokyo Institute of Technology, T. Sugiura, T. Kuga of the University of 
Tokyo, M. W. Jack and T. Hong of NTT Basic Research Laboratories for valuable discussions. 
%
%
%
\newcounter{q}
\setcounter{q}{118}

\end{document}